\documentclass[10pt,letterpaper]{article}
\usepackage{opex3}
\usepackage{cite}
\begin{document}

\title{Multi-wavelength holography with a single Spatial Light Modulator for ultracold atom experiments}

\author{David Bowman, Philip Ireland, Graham D. Bruce \\ and Donatella Cassettari$^{*}$}

\address{SUPA School of Physics and Astronomy, University of St Andrews, North Haugh,\\ St Andrews, KY16 9SS, UK}

\email{$^*$dc43@st-andrews.ac.uk} 

\begin{abstract}
We demonstrate a method to independently and arbitrarily tailor the spatial profile of light of multiple wavelengths and we show possible applications to ultracold atoms experiments. A single spatial light modulator is programmed to create a pattern containing multiple spatially separated structures in the Fourier plane when illuminated with a single wavelength. When the modulator is illuminated with overlapped laser beams of different wavelengths, the position of the structures is wavelength-dependent. Hence, by designing their separations appropriately, a desired overlap of different structures at different wavelengths is obtained. We employ regional phase calculation algorithms and demonstrate several possible experimental scenarios by generating light patterns with 670 nm, 780 nm and 1064 nm laser light which are accurate to the level of a few percent.  This technique is easily integrated into cold atom experiments, requiring little optical access.
\end{abstract}


\section{Introduction}

One of the most prevalent tools for the control of ultracold atoms is manipulation with light.  This manipulation can take many forms \cite{Cohen}, including: attractive or repulsive far-detuned trapping potentials; resonant light to image, cool, or pump the atoms to different electronic states; control of interactions via optical Feshbach resonances; molecule formation; manipulating the orbital angular momentum of the atoms; and imprinting phases and artificial gauge fields onto the atoms. Moreover, the introduction of high numerical-aperture imaging systems has increased the resolution with which atoms can be addressed with optical patterns \cite{Weitenberg}. Many of these applications can benefit from spatially-tailored light fields. 

With regard to atom trapping, in recent years there have been numerous approaches to tailoring the spatial profile of light, including acousto-optic deflection \cite{Henderson_09}, amplitude modulation \cite{Muldoon_12} and phase modulation \cite{Bergamini_04,Boyer_06,Franke-Arnold_07,Bruce_11ring,Bruce_11power,Gaunt_12,Nogrette_14,Bruce_14}. Phase modulation, which is used in this work, is the process of applying a spatially-varying retardation to the light field such that a target intensity pattern is realised in the Fourier plane (usually achieved by focussing the beam with a lens or a system of lenses). While all of these previous experiments have been conducted with one wavelength of light, there already exist proposals for cold atom experiments which make use of light fields with multiple wavelengths illuminating the atoms simultaneously but with different spatial distributions, e.g. \cite{Bernier,Sakhel,Uncu,Helm}.

Multi-wavelength holograms have previously been demonstrated for applications in full-colour display technology: by using multiple spatially-separated holograms \cite{Nakayama,Makowski2}; by time-division \cite{Shimobaba_03}, spatial-division \cite{Ito_04} or depth-division multiplexing \cite{Makowski}; and by illuminating a single phase-pattern with different wavelengths at appropriate angles \cite{Ito_04,Xue}. In this paper, we present a method which uses only a single hologram illuminated by co-propagating, overlapped laser beams. This is advantageous to a cold atoms experiment, where optical access is often limited. It also offers the flexibility of adding a new wavelength as the need arises with an easy alignment process and without additional computational demand. 

We tailor the spatial intensity distribution of up to three wavelengths (670~nm, 780~nm and 1064~nm) with a single retardation pattern applied across the whole active area of a Spatial Light Modulator (SLM). The required phase pattern is calculated using regional hologram-calculation algorithms \cite{pasienski, Harte}. We simultaneously illuminate this single hologram with different overlapped colours, creating a composite multi-wavelength pattern. Features that are spatially separated in the calculated intensity pattern are overlapped in the composite pattern, due to the position of the first order of diffraction scaling with wavelength.

In the following sections, we discuss the techniques required for the experimental realisation of light patterns suitable for ultracold atom experiments, by selecting example patterns of sub-diffraction-limited ring traps and a variety of larger-scale tailored potentials.

\section{Experimental setup}

In this work, holograms are generated using a single, reflective, phase-only SLM (Boulder~Nonlinear Systems P256) with $256 \times 256$ pixels which can impart a retardation between 0 and $2\pi$ in steps of $\pi / 128$. We illuminate the $6.14 \times 6.14$~mm SLM with laser beams generated by diode lasers at 670~nm, 780~nm and 1064~nm, each of which is independently expanded to a $1/e^{2}$ radius of 3.5~mm and overlapped using appropriate dichroic mirrors. After reflection from the SLM, the beams are focussed by a single off-the-shelf $f = 150$~mm achromatic doublet. The chromatic shift of the focal plane is measured to be less than 5~$\mu$m for our wavelengths. The undiffracted light (which contains $\sim50\%$ of the incident light power) from each of the three beams is overlapped in the Fourier plane, which is imaged onto a CCD camera. We program the retardation of the SLM to give optimal phase shift at the median wavelength, 780~nm, which we find to be a good compromise for all three wavelengths. 

\section{Multi-wavelength hologram design}

The calculation of a phase-only hologram for a desired intensity is a well-known inverse problem which requires numerical solutions in most cases. In cold atom experiments, only the region of the Fourier plane which is spatially close to the atoms is important, because the atoms interact only with light in this subset of the plane.  However, within this region of interest, the light field must be very smooth for trapping, in order to avoid unnecessary excitations. This requirement has been well captured by regional algorithms including the Mixed-Region Amplitude Freedom (MRAF) algorithm \cite{pasienski} and the conjugate gradient optimisation algorithm \cite{Harte}, which allow amplitude freedom in regions of the plane far from the atoms, lowering light-usage efficiency but increasing the accuracy and smoothness of the light within the region of interest. As shown in this paper, this regionality can also be exploited to diffract multiple wavelengths of light from a single applied phase.

\begin{figure}[ht]
\centering\includegraphics[width=0.75\textwidth]{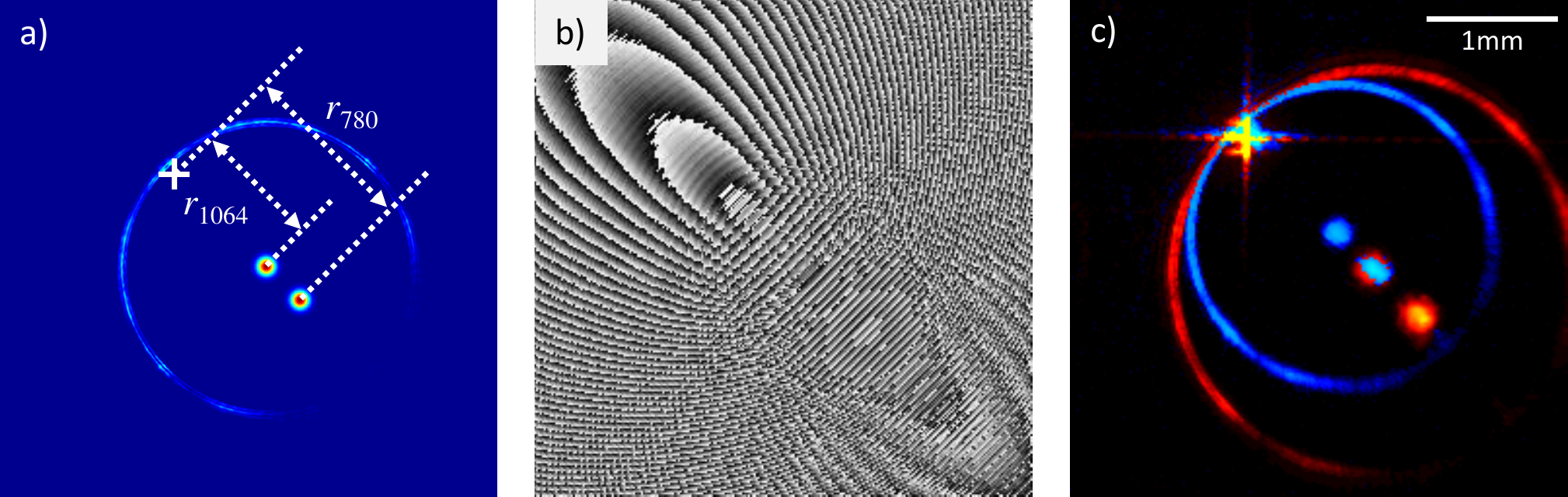}
\caption{
a) Target pattern containing Gaussians at $r_{1064}$ 
 and $r_{780}$ 
from the centre of the output plane, which is the location of undiffracted light in experimental light profiles. The ring-shaped feature is accumulation of light at the boundary of the region with amplitude-freedom, which is typical of the MRAF algorithm. 
b) Phase modulation required to achieve this target pattern.
c) Fourier-plane intensity acquired with 780~nm (blue) and 1064~nm (red) illumination. \label{Principle}}
\end{figure}

Target intensity patterns are programmed on a pixelated grid, where a pixel is chosen to be a fraction of a diffraction-limited spot in the output plane by appropriate use of input-plane padding \cite{pasienski}. The diffraction limit in the output plane is linearly dependent on the wavelength of the light, and therefore so are a target feature's size and position in the output plane. We use this fact to design target intensity distributions with a distinct feature for each wavelength of light with which the SLM will be illuminated. This is demonstrated by the simple pattern in Fig.~\ref{Principle}, which contains two Gaussians. The positions of the Gaussians in the target pattern are selected such that when the SLM is illuminated with 780~nm and 1064~nm light, the outer Gaussian at 780~nm is overlapped with the inner Gaussian at 1064~nm.
To achieve this, the Gaussians in the target are located at a distance $r_{1064}=208$~pixels (px) and $r_{780} = 1064 / 780 \times 208 = 283$~px from the centre of the plane. We calculate the required phase corresponding to monochromatic illumination of the SLM using the MRAF algorithm \cite{pasienski}.

When we illuminate the SLM with 780~nm light, we measure two Gaussians with $1/e^{2}$ waist of $91 \pm 3\mu$m, at $1.34\pm0.01$~mm and $1.93\pm0.01$~mm from the zeroth-order (undiffracted) light. With 1064~nm illumination, the Gaussians have $1/e^{2}$ waist of $130 \pm 9~\mu$m and are centred at $1.86 \pm 0.01$~mm and $2.57 \pm 0.01$~mm. The farther Gaussian at 780~nm and the nearer Gaussian at 1064~nm are therefore overlapped in the output plane, while the ratio of the Gaussian widths is equal to the ratio of the illuminating wavelengths, as expected. The presence of the additional spots of light which do not overlap reduces the light usage efficiency at each wavelength by a factor of two. 

\section{Sub-diffraction-limited optical trapping patterns}

\begin{figure}[ht]
\centering\includegraphics[width=1\columnwidth]{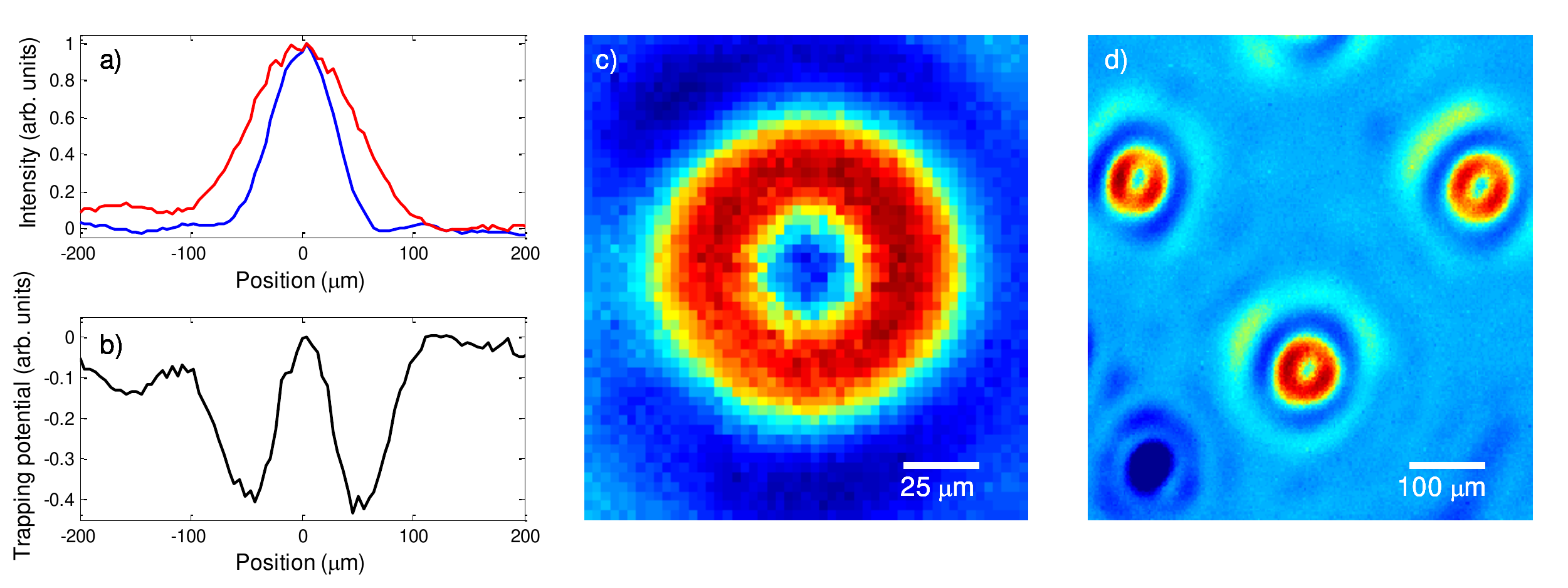}
\caption{a) Intensity profile of diffraction limited spots at 1064~nm (red) and 670~nm (blue). b) and c) Resulting trapping potential, calculated as the difference between the two intensity profiles. The thickness of the ring is $<50~\mu$m, below the diffraction limit of either colour. d) An array of three sub-diffraction-limited ring traps. \label{Ring}}
\end{figure}
One application of this multi-wavelength approach is overlapping two \emph{diffraction-limited} spots with wavelengths detuned either side of an atomic resonance. The red-detuned light gives an attractive trapping potential while the blue-detuned light gives a repulsive central potential, so that the resultant ring-shaped potential is below the diffraction limit of the optical system. While such a simple geometry can also be achieved without holography, an SLM can easily produce arrays of these small ring traps by tailoring the target profile, while further, more-complicated sub-diffraction-limited features can be designed. In Fig. \ref{Ring} we show examples by producing ring traps with 670~nm and 1064~nm light, which are respectively blue- and red-detuned from the main cooling transitions in rubidium. If we were to design ring traps with only 1064~nm light, their smallest attainable radius would be about twice the diffraction-limited spot radius at 1064~nm. By comparison, the radius of the two-colour ring is predicted to be a factor 3.6 smaller than the single-colour ring radius. Compared to a single-wavelength 1064 nm~ring, the thickness of the two-colour ring is again predicted to be reduced by a factor of 1.8, giving a higher trapping frequency. Experimentally, we de-magnify the beams impinging on the SLM to a $1 / e^2$ radius of $1.1$~mm to avoid clipping on the SLM aperture which would aberrate the point-spread function of the diffraction-limited spots. In order to acquire well-resolved images of the light pattern (see Fig. \ref{Ring}) we also magnify the Fourier plane by a factor of 2.5 using a confocal telescope. The Airy disks of the diffraction-limited spots have $1 / e^2$ radii of $53.9\pm0.5~\mu$m for 670~nm and $88\pm2~\mu$m for 1064~nm. When the ratio $I / \delta$ (where $I$ is the peak intensity and $\delta$ the detuning) is the same for both colours, we measure the resultant ring-shaped trapping potential to have a radius of $53.5\pm0.5~\mu$m and a $1 / e^2$ half-width of $44\pm3~\mu$m, consistent with our predictions. The generation of arrays of these ring traps, as shown in Fig. \ref{Ring}d), is accomplished by designing a target profile containing two arrays, with array periodicities and placements being defined by the wavelengths of the illuminating light.

\section{Multi-wavelength feedback algorithm}

Experimental realisations of holograms larger than the diffraction limit are subject to aberrations which can be overcome with a feedback algorithm such as that described in \cite{Bruce_14}. In this work we regionalise this algorithm to correct only the relevant output-plane feature for each wavelength. 

As an example, we consider dressing a standing-wave optical lattice with the trapping potential proposed in \cite{Bernier} for entropy removal of lattice-confined fermionic atoms, which consists of a central attractive dimple potential surrounded by a repulsive barrier to separate the dimple from the remainder of the ensemble.  In this scheme, which could be implemented in state-of-the-art quantum gas microscopes \cite{Haller}, the dimple and repulsive barrier produce a low-entropy region in the centre, and the high-entropy atoms outside the repulsive barrier can be removed to lower the entropy per particle.  This removal could, for example, be achieved using the single-site addressing techniques of Weitenberg, \emph{et al} \cite{Weitenberg}.  In brief, a magic-wavelength light beam is used to impart a differential light shift of the hyperfine ground states of the atoms by $\sim100$kHz. The light-shifted atoms are transferred between the ground states using a microwave pulse. All atoms in one or other state can now be selectively removed by resonant excitation with high efficiency while atoms in the other state remain in the ground state of the potential well \cite{Weitenberg}.

The spatially-varying light patterns in this method can be generated with a single SLM: the central attractive-dimple potential is calculated using 1064~nm light and the repulsive barrier to separate the dimple from the remainder of the ensemble is generated using 670~nm light (which is blue-detuned from resonance for $^{40}$K). In addition, we design a tailored profile for magic-wavelength addressing of atoms outside the trapping potential, using our 780~nm light as proof-of-principle (the magic-wavelength of $^{40}$K is $768.4$~nm and therefore within easy reach of our method). In order to achieve the desired differential light shift between the $4S_{1/2}$ ground states of $^{40}$K, 10~mW in the magic-wavelength beam could form, for example, a ring of diameter 120~$\mu$m with 50$~\mu$m width. The differential light shift  from the 1064~nm light will be very small ($\sim40$~Hz for 1~W of power and a trap radius of $\sim50~\mu$m).

\begin{figure}[htbp]
\centering\includegraphics[width=0.3\columnwidth]{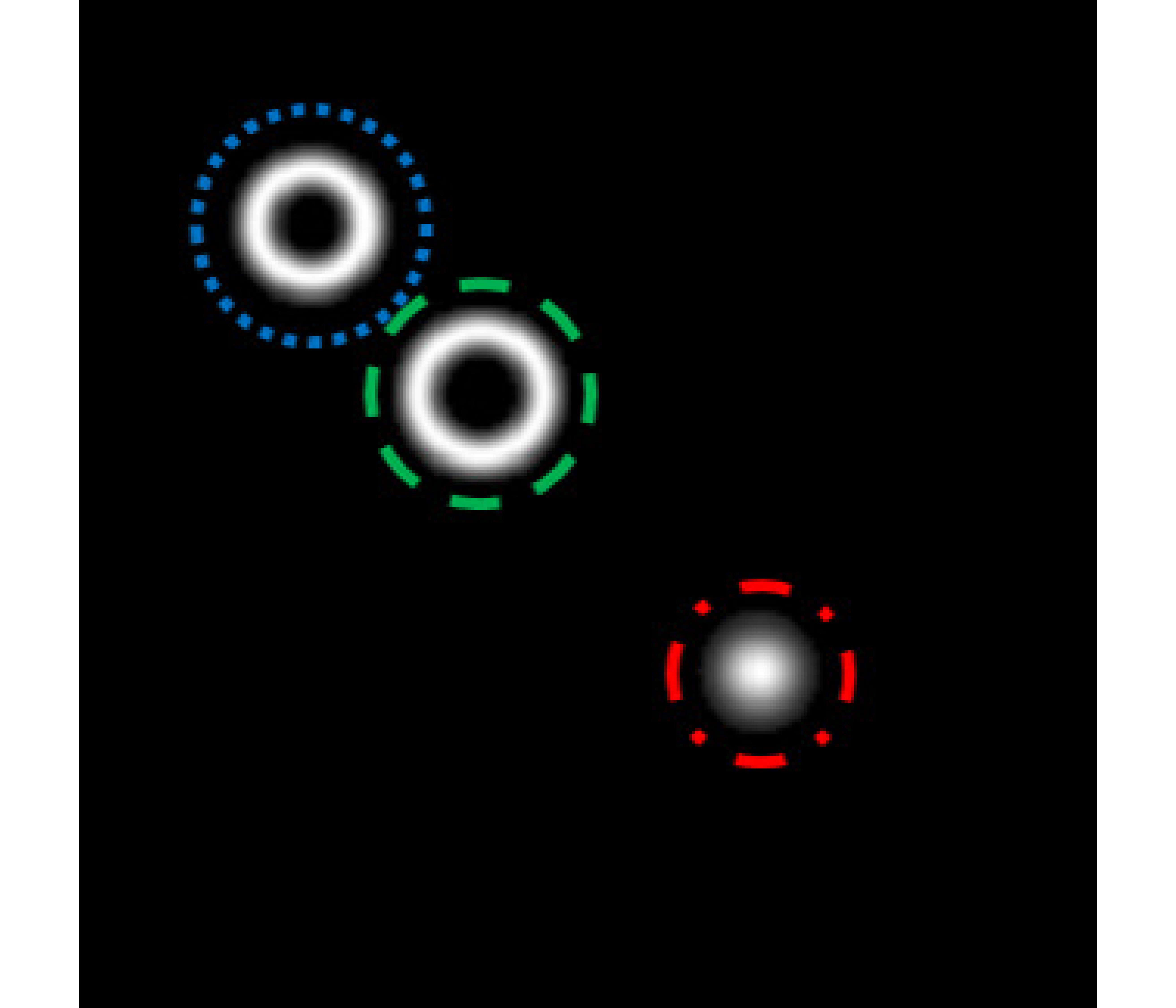}
\caption{Target pattern for lattice-based entropy reduction, showing the subset of the plane containing the measure regions to be used within the feedback algorithm. The target consists of two rings and a Gaussian, designed for 670~nm, 780~nm and 1064~nm. The blue, green and red regions will be optimised for 670~nm, 780~nm and 1064~nm respectively. \label{Regions}}
\end{figure}

\begin{figure}[htbp]
\centering\includegraphics[width=0.74\columnwidth]{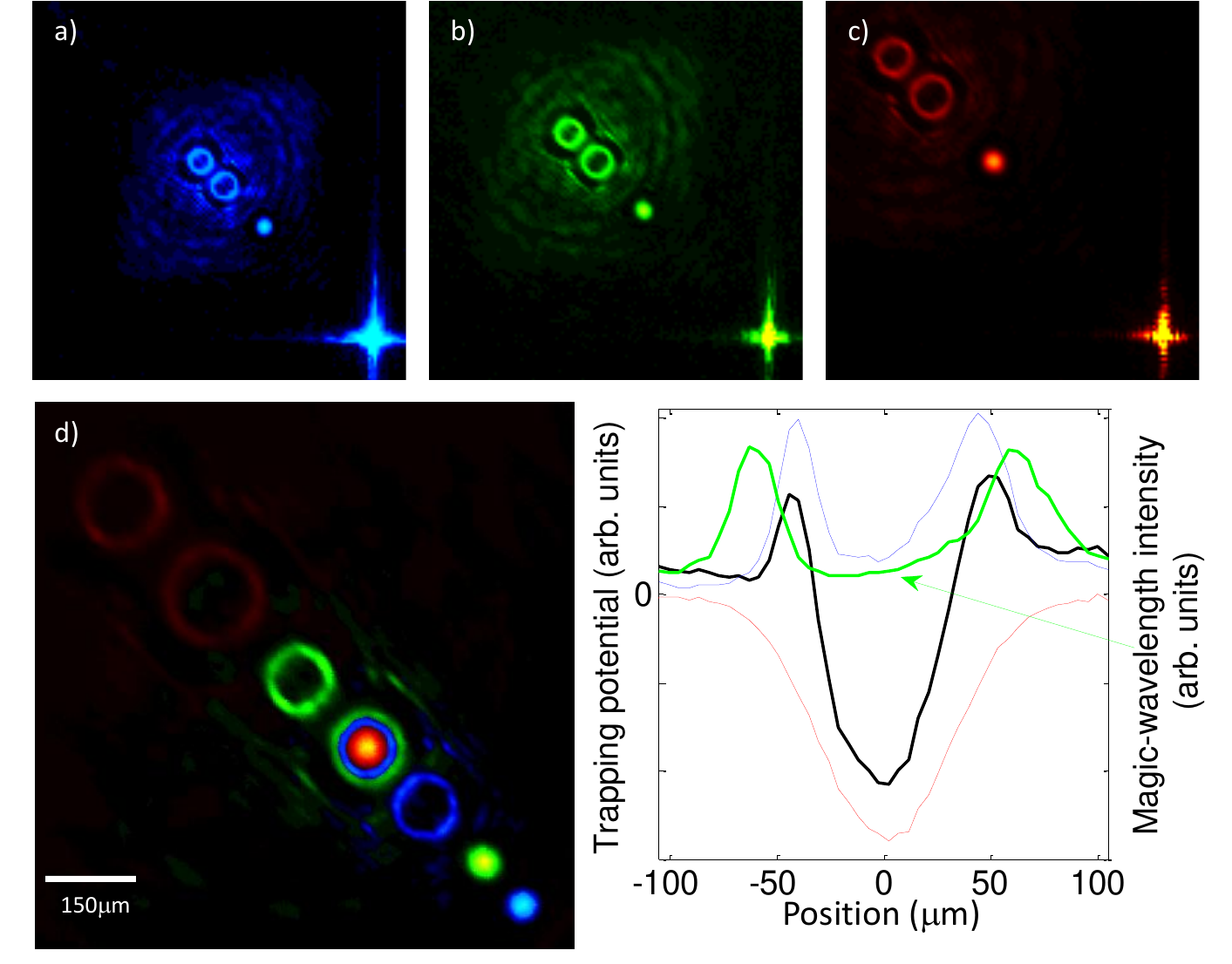}
\caption{Entropy-separation light pattern illuminated with a)~670~nm, b)~780~nm, c)~1064~nm, d) all three wavelengths. e) Region where the three wavelengths overlap: trap profile versus position for 670~nm (blue) and 1064~nm (red) and the resultant combined trapping potential (black). The magic-wavelength light (green) allows spatially-selective microwave state transfer of atoms outside the resultant potential, after which a resonant light pulse (which does not need to be spatially controlled and may be directed orthogonally to the SLM-generated pattern) can selectively remove the transferred atoms.
\label{FeedbackImg}}
\end{figure}

Our target intensity profile, shown in Fig. \ref{Regions}, therefore has three components: a Gaussian with $1/e^{2}$ radius of 12.4~px at $r=126$~px from the plane centre; a 27.4~px diameter Gaussian ring with $1/e^{2}$ radial half-width of 5.7~px at $r=172$~px; and a second Gaussian ring of 24~px diameter and 5.7~px radial half-width at $r=200$~px. For each of these features we assign a measure region as shown in Fig. \ref{Regions}, which is where the feedback algorithm compensates aberrations. Converting to output-plane coordinates, all three features will be centred at the same coordinate and their relative sizes will be such that the inner diameter of the 670~nm ring overlaps with the edge of the 1064~nm Gaussian and the outer diameter of the 670~nm ring overlaps with the inner diameter of the 780~nm ring. We calculate an initial phase using the MRAF algorithm, which is displayed on the SLM and illuminated with each of the three lasers in turn. To implement the feedback algorithm, the acquired image is compared to the target profile as in \cite{Bruce_14}. However, here we compare the acquired image to the target within only the appropriate region for each wavelength, e.g. we ignore the two rings in the image acquired with 1064~nm. Within this region the difference between target and acquired pattern is added to the original target to create a new target for a subsequent iteration of MRAF. For all subsequent iterations the same routine applies, except that the difference between target and acquired pattern is added to the \emph{previous} iteration's target. For the example shown, four iterations of feedback reduce the rms error of the 670~nm ring to $8.1\%$, the 780~nm ring to $9.3\%$ and the 1064~nm Gaussian to $0.5\%$, from $21.3\%$, $20.9\%$ and $8.6\%$ respectively. While the algorithm does leave some magic-wavelength light within the region of the central dimple, this will not have a deleterious effect as it will serve to light shift atoms in this region farther from resonance with the microwave source. These feedback-enhanced patterns are shown in Fig. \ref{FeedbackImg} together with the resulting composite pattern, which could be superimposed on fermions trapped in an optical lattice. If necessary, the light in the remainder of the output plane can be blocked with a pinhole.

\begin{figure}[htbp]
\centering\includegraphics[width=0.74\columnwidth]{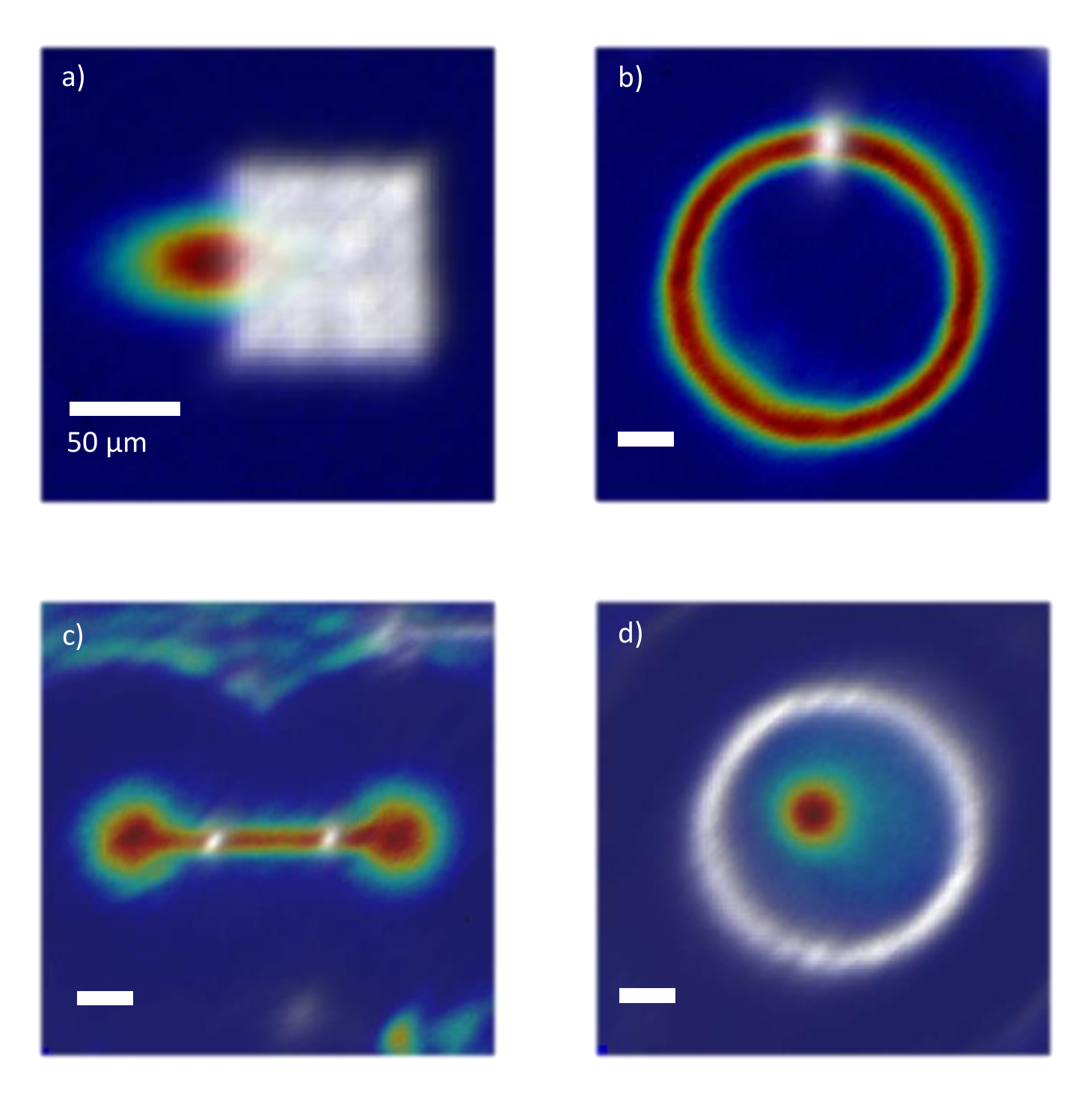}
\caption{A selection of light patterns suitable for cold atom experiments, generated with 1064~nm (colour) and 780~nm (gray-scale, shown with background transparency for clarity). The scale bar in each image denotes $50~\mu$m. a) Elliptical red-detuned trap, partially illuminated by off-resonant light with a sharp edge for phase manipulation of Bose--Einstein condensates. b) Red-detuned ring trap with blue-detuned barrier. c) Red-detuned double well, connected by a thin channel interrupted by blue-detuned barriers. d) Red-detuned trap with additional offset attractive dimple, bounded by a blue-detuned ring to create hard walls to the trapping potential.
\label{MorePatterns}}
\end{figure}

To show the generality of our approach, we have generated a range of further light patterns which may have use in ultracold atom experiments, as shown in Fig. \ref{MorePatterns}. Pattern a) consists of a red-detuned elliptical optical trap, which is partially superimposed with a 6th-order super-Lorentzian profile detuned by several tens of GHz from an atomic resonance. This second beam causes the phase of the illuminated and non-illuminated sections of the condensate to evolve at different rates, causing a phase-slip which leads to soliton formation \cite{Becker}. Our feedback algorithm gives a smooth ellipse with $1.4\%$ rms error and a super-Lorentzian with $1.7\%$ rms variation in the flat-top. Fig. \ref{MorePatterns}b) shows a red-detuned ring trap with an independent blue-detuned barrier to be used in soliton-interferometry as proposed in \cite{Helm}. The rms variation around the circumference of the ring is $4.5\%$ while the barrier rms error is $4.3\%$. In previous work \cite{Bruce_11ring} we have demonstrated that ring traps of this accuracy are adequate for cold atom experiments. 

The study of conduction between two reservoirs along a one-dimensional channel interrupted by repulsive barriers was recently proposed in \cite{Simpson}. An optical potential in this geometry is shown in Fig. \ref{MorePatterns}c), where the rms variation along the conduction channel is $3.3\%$. Finally, the authors of \cite{Sakhel} have proposed experiments in attractive potentials with a narrower dimple which is offset from the centre of the trap, with the potential bounded by hard walls. The blue-detuned ring in Fig. \ref{MorePatterns}d) will produce a steep potential at the edges of the red-detuned trap, which has an rms error of $1.3\%$. 

A test of the success of our approach to multi-wavelength hologram generation is the control of the relative positions of features at different wavelengths. We compare measured and designed multi-wavelength patterns by assuming that one wavelength is well-aligned to its target, and finding the offset of patterns in other wavelengths. When the illuminating beams are well-overlapped both before the SLM and in the zeroth-order of SLM diffraction, after feedback we find that the difference between target and measured positions for the second (and third) wavelengths is less than $5~\mu$m (one camera pixel) irrespective of the pattern size and distance from the zeroth order.

\section{Conclusion}

We have presented a simple method to produce accurate, multi-wavelength light patterns for use in cold atom experiments. This method exploits the regionality of hologram-calculation algorithms to calculate a single phase pattern which can be illuminated with overlapped beams. In particular, illumination with overlapped beams is desirable from the perspective of cold atoms experiments which often have limited optical access into the experimental vacuum chamber. The method also offers the flexibility to add another wavelength to the experiment with a simple alignment procedure.

While the chromatic shift of the focal plane of our optical system is small, for more advanced optical systems such as those in \cite{Weitenberg,Haller}, the chromatic shift between the wavelengths may be larger. This may be overcome by independently adjusting the initial collimation of each of the laser beams before they impinge on the SLM. An improvement in the accuracy of the holograms may be achieved by using an SLM with extended phase-modulation, as in \cite{Jesacher}, while an SLM with a higher pixel resolution would allow for more detailed light patterns.

\section*{Acknowledgments}

The authors acknowledge funding from the Leverhulme Trust RPG-2013-074 and from EPSRC.

\end{document}